\documentclass[sn-mathphys,Numbered]{sn-jnl}

\usepackage{graphicx}%
\usepackage{multirow}%
\usepackage{amsmath,amssymb,amsfonts}%
\usepackage{amsthm}%
\usepackage{mathrsfs}%
\usepackage[title]{appendix}%
\usepackage{xcolor}%
\usepackage{textcomp}%
\usepackage{manyfoot}%
\usepackage{booktabs}%
\usepackage{algorithm}%
\usepackage{algorithmicx}%
\usepackage{algpseudocode}%
\usepackage{listings}%

\usepackage{makecell}%

\begin{document}

\title{From CNN to Transformer: A Review of Medical Image Segmentation Models}


\author[1]{\fnm{Wenjian} \sur{Yao}}\email{1304319616@qq.com}

\author[1]{\fnm{Jiajun} \sur{Bai}}\email{haku\_bill@outlook.com}

\author[2]{\fnm{Wei} \sur{Liao}}\email{770147659@qq.com}

\author[1]{\fnm{Yuheng} \sur{Chen}}\email{823368720@qq.com}

\author*[1]{\fnm{Mengjuan} \sur{Liu}}\email{mjliu@uestc.edu.cn}

\author*[3, 4]{\fnm{Yao} \sur{Xie}}\email{xieyao@med.uestc.edu.cn}

\affil[1]{Network and Data Security Key Laboratory of Sichuan Province, University 
 of Electronic Science and Technology of China, Chengdu, 610054, China}

\affil[2]{Department of obstetrics and gynaecology, Deyang People's Hospital, Deyang 618000,China}

\affil[3]{Department of obstetrics and gynaecology, Sichuan Provincial People's Hospital, University of Electronic Science and Technology of China, Chengdu, China}

\affil[4]{Chinese Academy of Sciences Sichuan Translational Medicine Research Hospital, Chengdu 610072, China}


\abstract{Medical image segmentation is an important step in medical image analysis, especially as a crucial prerequisite for efficient disease diagnosis and treatment. The use of deep learning for image segmentation has become a prevalent trend. The widely adopted approach currently is U-Net and its variants. Additionally, with the remarkable success of pre-trained models in natural language processing tasks, transformer-based models like TransUNet have achieved desirable performance on multiple medical image segmentation datasets. In this paper, we conduct a survey of the most representative four medical image segmentation models in recent years. We theoretically analyze the characteristics of these models and quantitatively evaluate their performance on two benchmark datasets (i.e., Tuberculosis Chest X-rays and ovarian tumors). Finally, we discuss the main challenges and future trends in medical image segmentation. Our work can assist researchers in the related field to quickly establish medical segmentation models tailored to specific regions.}




\maketitle

\section{Introduction}\label{sec1}

With the continuous development of medical imaging technology, medical images have become essential for disease diagnosis and treatment planning \cite{2016Computer}. Medical image segmentation plays a vital role among the foundational and critical techniques in medical image analysis. Medical image segmentation refers to the identification of organ or lesion pixels from medical images such as CT or MRI \cite{2016Lung} \cite{2017Automatic}. It is one of the most challenging tasks in medical image analysis, aiming to convey and extract crucial information about the shape and volume of these organs or tissues. Traditional methods for medical image segmentation primarily rely on manual feature extraction by physicians or handcrafted designs based on image processing techniques and mathematical models, such as thresholding \cite{otsu1979threshold}, edge detection \cite{Magnier2017Edge}, and morphological operations. These methods offer a certain level of interpretability and controllability. However, due to the complexity and diversity of medical images, as well as the specificity of medical image segmentation tasks, traditional segmentation methods have certain limitations. Handcrafted algorithms fail to meet the requirements of efficiency and accuracy when dealing with a large number of medical images for segmentation tasks. Moreover, manual feature extraction from medical images requires physicians with rich expertise and experience, making them susceptible to subjective factors. 

Deep learning techniques have been widely applied in medical image segmentation in recent years to address the issues above. Through deep feature learning, models can extract semantic information from images, thereby improving segmentation accuracy and flexibly adapting to different medical image datasets and tasks. Segmentation models based on convolutional neural networks (CNNs) have achieved remarkable results. For example, the U-Net model \cite{ronneberger2015u} won first place in the ISBI 2015 Cell Segmentation Challenge, and the SegNet \cite{badrinarayanan2017segnet} model demonstrated good performance in semantic segmentation tasks on the CamVid dataset, among others. However, convolutional neural networks have limited modeling capabilities for long-range dependencies, making it challenging to exploit the semantic information within images fully. 

Recently, some new segmentation models have been proposed, including TransUNet \cite{chen2021transunet} and Swin-Unet \cite{cao2022swin}. Trans-UNet is a segmentation model that introduces Transformer modules \cite{carion2020end} to improve the model's ability to model long-range dependencies. The Transformer module adopts a self-attention mechanism, which calculates the similarity between each position and other positions in the input sequence, resulting in a weight vector. This weight vector is used to compute weighted representations for each position, facilitating the interaction and integration of global information. In other words, the Transformer model can effectively capture the correlations between different positions in the input sequence through the self-attention mechanism, thereby better understanding and processing sequential data. In Trans-Unet, the Transformer module is embedded within a U-shaped architecture to extract global information from the image, enhancing the model’s semantic representation capability and making it more suitable for handling large-sized, high-resolution medical images.

Swin-Unet, on the other hand, is another novel segmentation model that introduces the Swin Transformer module \cite{liu2021swin} to improve computational efficiency. The Swin Transformer is a hierarchical self-attention mechanism that decomposes the input feature map into multiple small patches, with each patch independently computing attention weights, thus reducing computational complexity. The Swin Transformer module in Swin-Unet is combined with a U-shaped architecture, allowing for the extraction of global information from the image while reducing computational complexity and memory consumption. This makes it more suitable for medical image segmentation tasks.

Despite the vibrant development of medical image segmentation techniques in recent years, there is still a lack of comprehensive review papers on the application of deep learning models in medical image segmentation, particularly the introduction of the latest segmentation models and quantitative performance comparisons among these models. The literature \cite{aljuaid2022survey} \cite{abdou2022literature} only covers traditional CNN-based segmentation models, while literature \cite{asgari2021deep} focuses solely on the model structure without quantitative evaluation.

This paper conducts a survey on the four most representative medical image segmentation models in recent years: U-Net, UNet++ \cite{zhou2018unet++}, TransUNet, and Swin-Unet. The characteristics of these models are analyzed theoretically, and their performance is quantitatively evaluated on two benchmark datasets. Finally, we discuss the main challenges and future development trends in medical image segmentation. Furthermore, we have shared all experimental source code and detailed model configuration parameters on GitHub to assist related researchers in quickly understanding these models and modeling new segmentation tasks. 

The rest of this paper is organized as follows: Section 2 describes representative medical image segmentation methods. Section 3 introduces the datasets and provides relevant experimental details. Section 4 presents the evaluation results of each model conducted in our study. Finally, the paper concludes with an outlook on the challenges and future developments in medical image segmentation.

\section{Typical Medical Image Segmentation Models}\label{sec2}

In recent years, medical image segmentation has made great progress with the help of deep learning \cite{ker2017deep}. Convolutional neural networks (CNNs) \cite{krizhevsky2012imagenet}, especially fully convolutional networks (FCNs) \cite{long2015fully}, dominate medical image segmentation. With the development in medical image segmentation, among different model variants, U-Net has become the de facto choice, consisting of a symmetric encoder-decoder network with jump-over connections for enhanced detail retention. Based on this neural network, image features can be automatically extracted and used in segmentation tasks. Several deep learning models have been used with excellent results in medical image segmentation, such as U-Net, UNet++, 3D U-Net \cite{cciccek20163d}, V-Net \cite{milletari2016v}, Attention-UNet \cite{schlemper2019attention}, TransUNet and Swin-Unet.

\subsection{U-Net}\label{subsec1}

\begin{figure}[htbp]
\centering
\includegraphics[width=.9\textwidth]{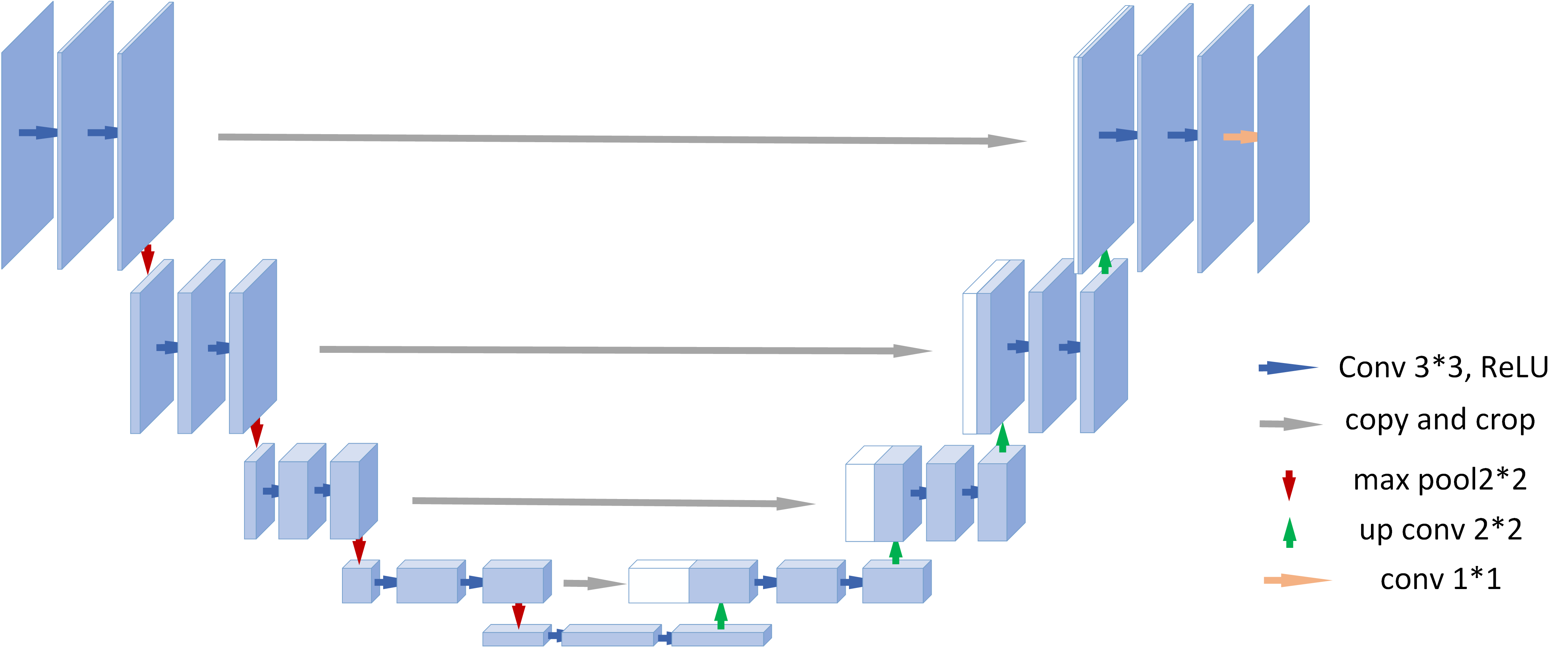}
\caption{Structure of U-Net}
\label{fig:fig1}
\end{figure}

U-Net is one of the most well-known network architectures in the medical image segmentation models. It was proposed by Ronneberger et al. \cite{ronneberger2015u} for the ISBI Challenge in 2015. The U-Net model is considered a classic model in medical image segmentation and has been widely applied to various tasks, including CT, MRI, and X-ray segmentation. The model structure is depicted in Fig. \ref{fig:fig1}. Its success lies in combining the deep feature extraction capability of convolutional neural networks (CNNs) with the pixel-level segmentation ability of fully convolutional networks (FCNs). It also incorporates techniques such as skip connections to leverage both low-level and high-level feature information, thereby improving segmentation accuracy and robustness. 

The U-Net network consists of a contracting path and an expanding path. The contracting path follows a typical architecture of convolutional networks. In each downsampling step, the number of feature channels is doubled. Each step in the expanding path includes upsampling feature maps, halving the number of feature channels, and concatenating them with the correspondingly cropped feature maps from the contracting path. In the final layer, a 1x1 convolution is applied to map each 64-component feature vector to the desired number of classes. The network consists of a total of 23 convolutional layers. 

Since the introduction of the U-Net model, several improved versions based on U-Net have emerged, including UNet++, Attention-UNet, TransUNet, and Swin-Unet, among others. These models build upon the advantages of the original U-Net model and further enhance segmentation performance by introducing attention mechanisms, transformational network structures, and other techniques. As a result, the U-Net model holds an important position and influence in medical image segmentation. 

\subsection{UNet++}\label{subsec2}

\begin{figure}[htbp]
\centering
\includegraphics[width=.9\textwidth]{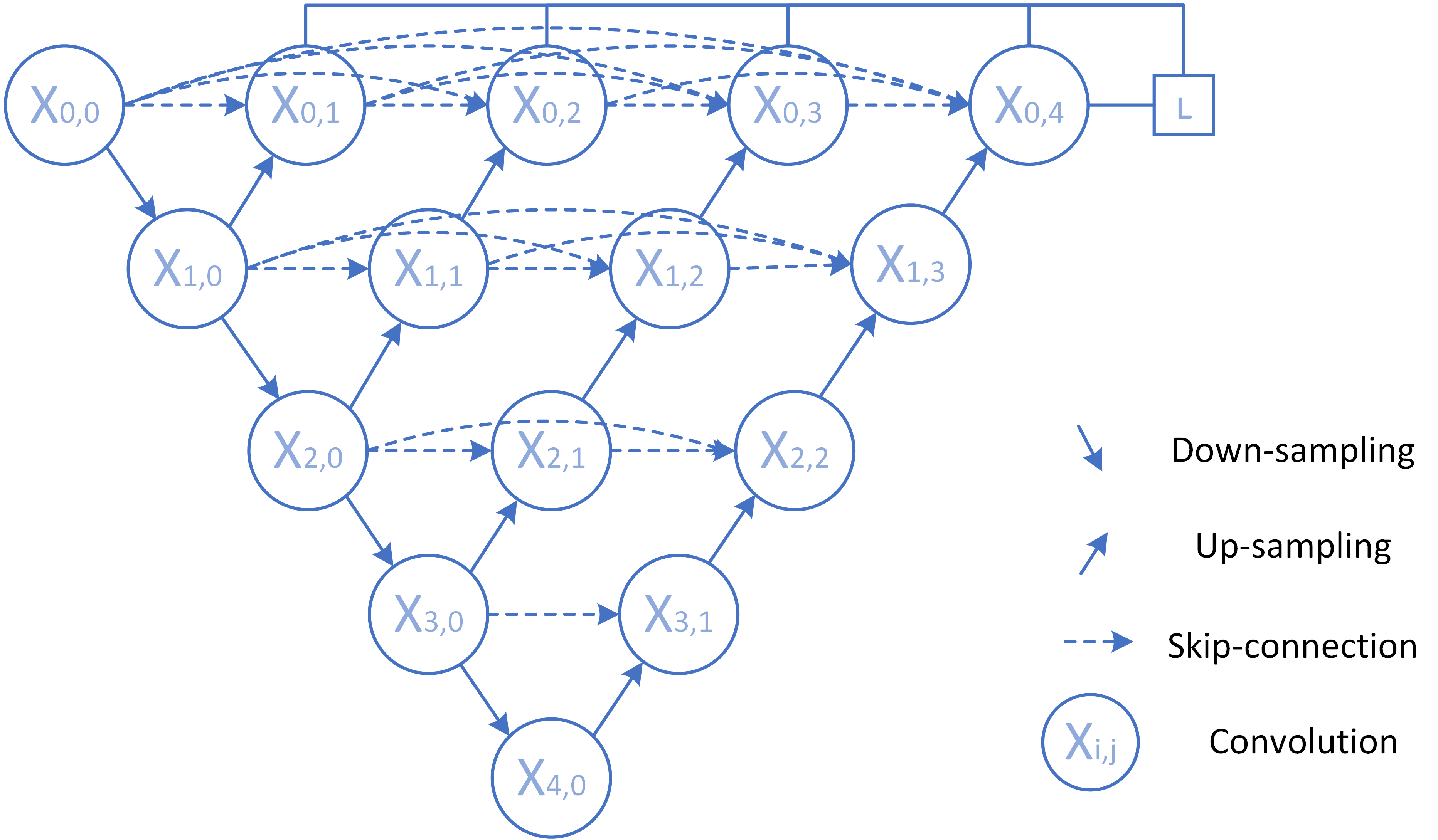}
\caption{Structure of UNet++}
\label{fig:fig2}
\end{figure}

The UNet++ network architecture was proposed by Zhou et al. \cite{zhou2018unet++} in 2018, introducing the concept of dense connections to the U-Net network. The model structure is depicted in Fig. \ref{fig:fig2}. UNet++ builds upon the U-Net model while retaining the long skip connections and adds more short-skip connection paths and upsampling convolution blocks to form new levels of encoders. The U-shaped connectivity structure in UNet++ is achieved by fusing each encoder in the decoder with other encoders at the same level. Specifically, each encoder receives feature maps of the same scale from other encoders and concatenates them to obtain more discriminative feature representations. Furthermore, the later proposed attention-UNet++ \cite{li2020attention} improves the feature map concatenation by adding attention mechanisms to enhance the focus and extraction of important features during encoder fusion. 

UNet++ captures features from different levels by introducing dense connections, enabling the extraction of feature information from different layers and scales. These features are integrated into the final prediction to improve segmentation accuracy. The idea of dense connections originated from DenseNet \cite{huang2017densely}. Prior to DenseNet, the evolution of convolutional neural networks typically involved increasing the depth or width of the networks. DenseNet introduced a new structure by reusing features, which not only alleviated the problem of gradient vanishing but also reduced the number of model parameters. In the original U-Net network architecture, the use of depth supervision in the intermediate hidden layers addresses the issue of gradient vanishing during UNet++ training. It also allows for network pruning during the testing phase, reducing the inference time of the model. 

\subsection{TransUNet}\label{subsec3}

\begin{figure}[htbp]
\centering
\includegraphics[width=.9\textwidth]{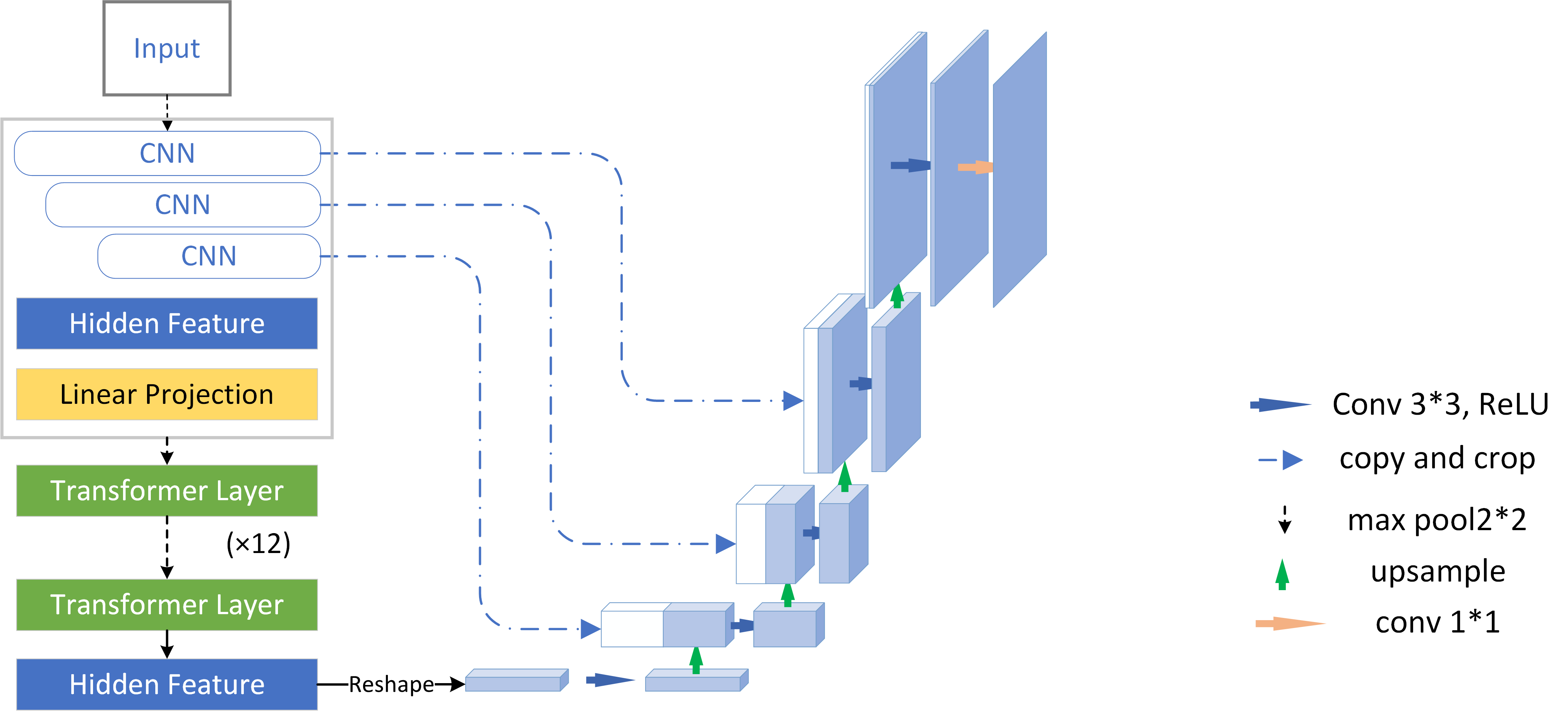}
\caption{Structure of TransUNet}
\label{fig:fig3}
\end{figure}

TransUNet network architecture was proposed by Chen et al. \cite{chen2021transunet} in 2021 as a Transformer-based segmentation network. The model structure is depicted in Fig. \ref{fig:fig3}. TransUNet builds upon the U-Net model by introducing a hybrid encoder that combines CNN and Transformer to address the limitations of traditional convolutional neural networks in modeling long-range dependencies and handling large-sized images. The core of TransUNet is the Transformer module \cite{carion2020end}, which consists of multi-head self-attention mechanisms and feed-forward neural networks. The multi-head self-attention mechanism captures dependencies between different positions in the image, establishing global contextual information in the feature representation. This enables Trans-UNet to handle long-range dependencies better, capture semantic information in the image, and improve the model's representation capacity and generalization performance.

Specifically, TransUNet first uses CNN to extract features and generate feature maps of the input image. These feature maps are then divided into patches of size 1x1 and fed into an additional stack of 12 Transformer modules. This hybrid structure combines convolutional neural networks' feature extraction capability with effective global information modeling using Transformer modules, yielding better performance than using pure Transformers as encoders. The decoder in TransUNet performs upsampling on the encoded features and combines them with high-resolution CNN feature maps to enrich the semantic information, achieving more precise localization. The final step involves restoring the feature maps to the original image size and generating pixel-level segmentation results. Compared to the traditional U-shaped models that use convolutional neural networks, TransUNet introduces a stack of 12 Transformer modules, significantly increasing the number of parameters. It increases the difficulty of model training. In this study, a suboptimal approach of reducing the batch size was employed to meet the training requirements of TransUNet on GPUs. 

\subsection{Swin-Unet}\label{subsec4}

\begin{figure}[htbp]
\centering
\includegraphics[width=.9\textwidth]{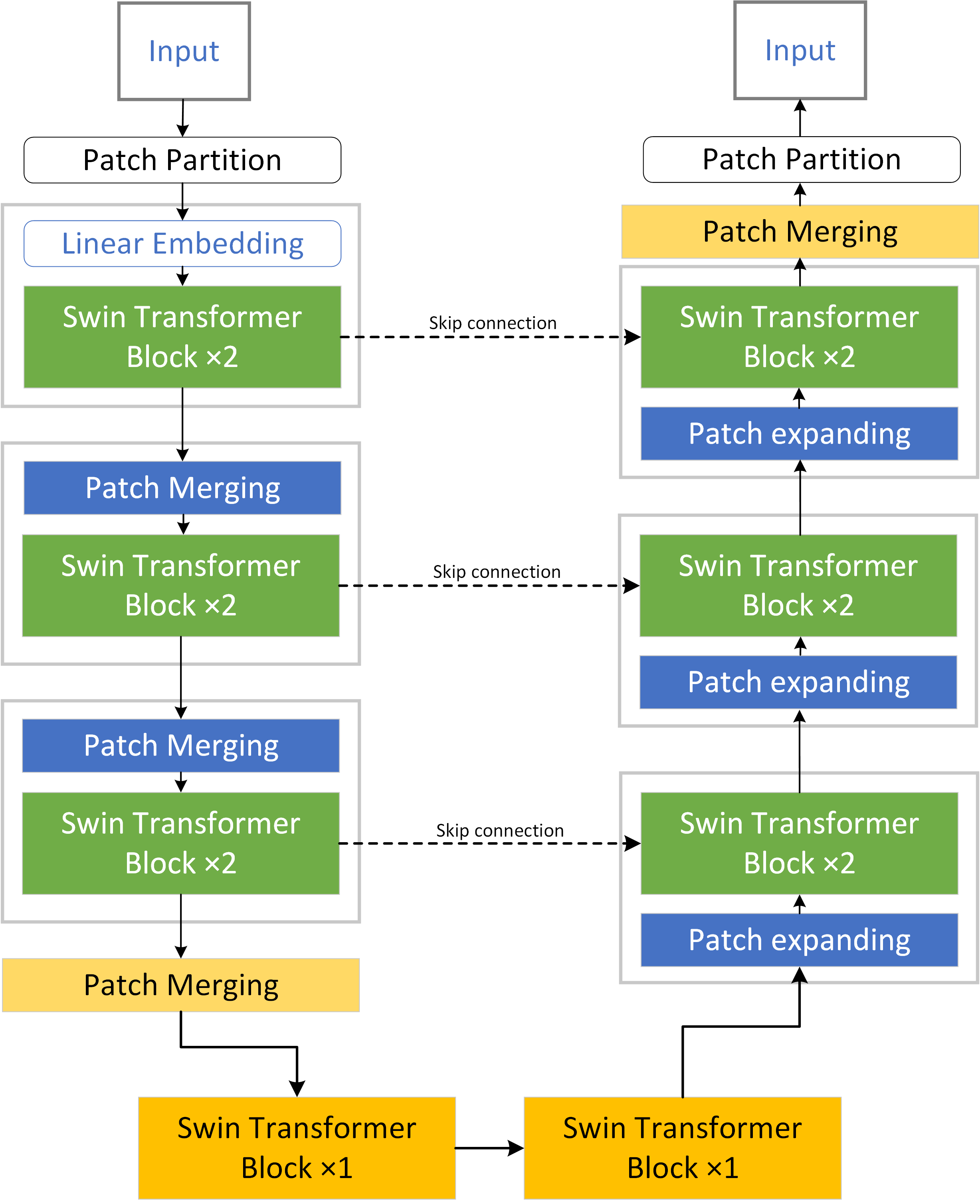}
\caption{Structure of Swin-Unet}
\label{fig:fig4}
\end{figure}

Swin-Unet network architecture was proposed by Cao et al. \cite{cao2022swin} in 2023. The model structure is depicted in Fig. \ref{fig:fig4}. Unlike Trans-Unet, which replaces the convolutional blocks in the U-Net encoder with Transformer blocks, Swin-Unet utilizes Swin Transformer blocks \cite{liu2021swin} to extract hierarchical features from the input image. Swin-Unet is the first purely Transformer-based U-shaped architecture. Swin Transformer extends the traditional Transformer's one-dimensional sequence to two-dimensional image blocks and adopts a hierarchical attention mechanism to capture features over a larger receptive field. This structure is similar to the hierarchical structure in convolutional neural networks and serves feature extraction. Additionally, Swin Transformer introduces the mechanism of shifting windows on top of the self-attention mechanism. By limiting the attention calculation to windows in the vicinity of the current region, Swin-Unet better preserves positional information and further improves the model's performance.  

In Swin-Unet, Swin Transformer is applied in the encoding, bottleneck, and decoding modules. Importantly, the compression of each layer's features in Swin-Unet is smaller than in TransUNet. Instead of adding additional Transformer modules, Swin-Unet replaces the convolutional modules with Transformer modules, effectively reducing the number of model parameters. Overall, Swin-Unet leverages the advantages of Swin Transformer and U-Net to provide a promising approach for medical image segmentation. It has demonstrated competitive performance in various segmentation challenges and benchmarks. 

\section{Experimental Setup}
\subsection{Datasets}\label{subsec31}

This subsection presents the datasets used for evaluating the four typical segmentation models. 

\begin{figure}[htbp]
	\centering
	\begin{minipage}{0.19\linewidth}
		\centering
		\includegraphics[width=0.9\linewidth]{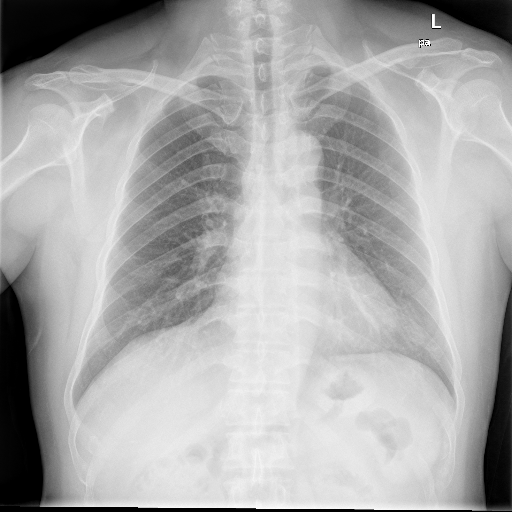}
	\end{minipage}
	\begin{minipage}{0.19\linewidth}
		\centering
		\includegraphics[width=0.9\linewidth]{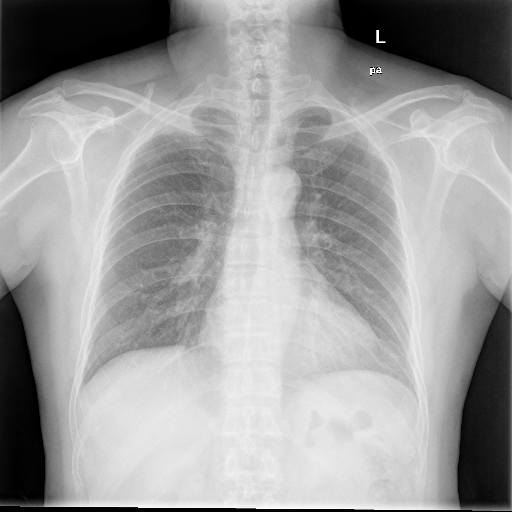}
	\end{minipage}
	\begin{minipage}{0.19\linewidth}
		\centering
		\includegraphics[width=0.9\linewidth]{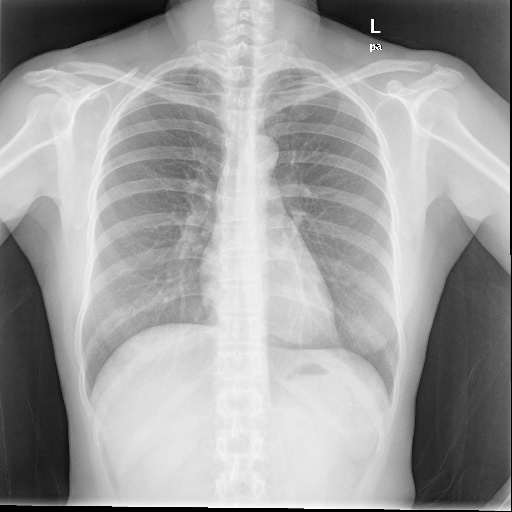}
	\end{minipage}
	\begin{minipage}{0.19\linewidth}
		\centering
		\includegraphics[width=0.9\linewidth]{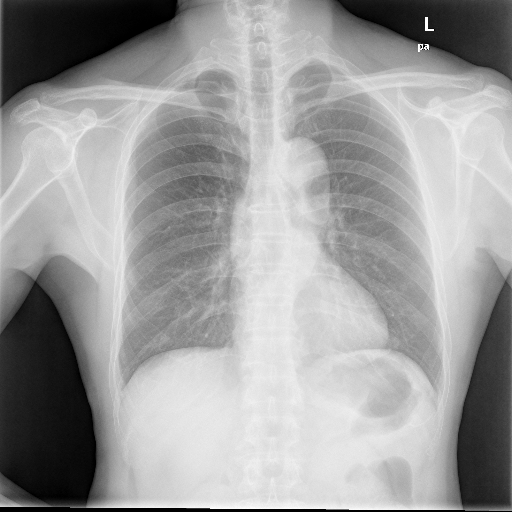}
	\end{minipage}
 
        \vspace{0.2cm}
        
	\begin{minipage}{0.19\linewidth}
		\centering
		\includegraphics[width=0.9\linewidth]{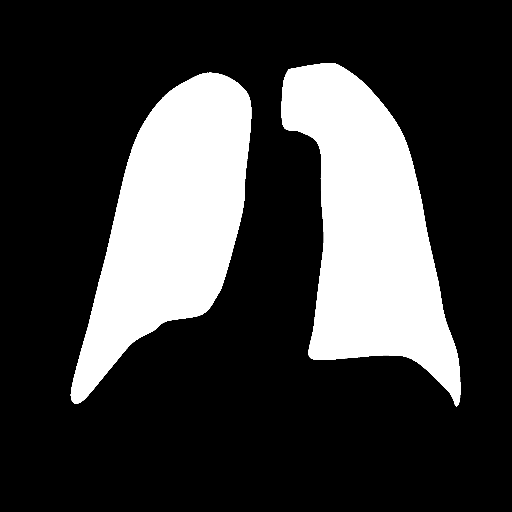}
	\end{minipage}
	\begin{minipage}{0.19\linewidth}
		\centering
		\includegraphics[width=0.9\linewidth]{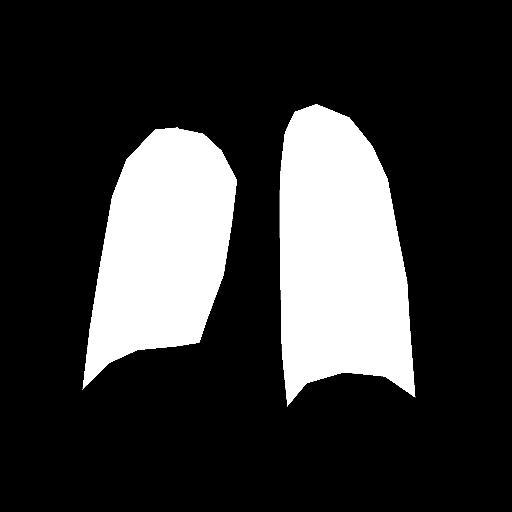}
	\end{minipage}
	\begin{minipage}{0.19\linewidth}
		\centering
		\includegraphics[width=0.9\linewidth]{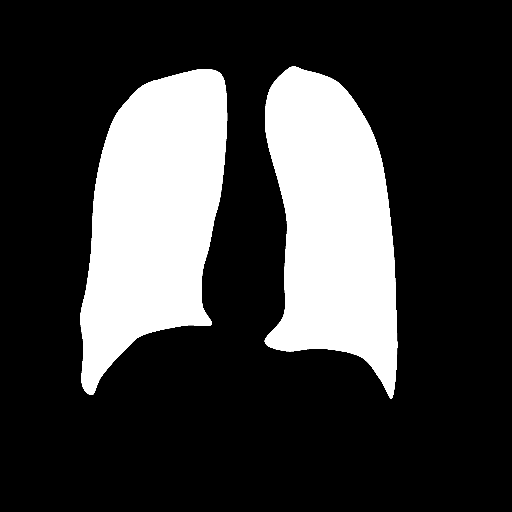}
	\end{minipage}
	\begin{minipage}{0.19\linewidth}
		\centering
		\includegraphics[width=0.9\linewidth]{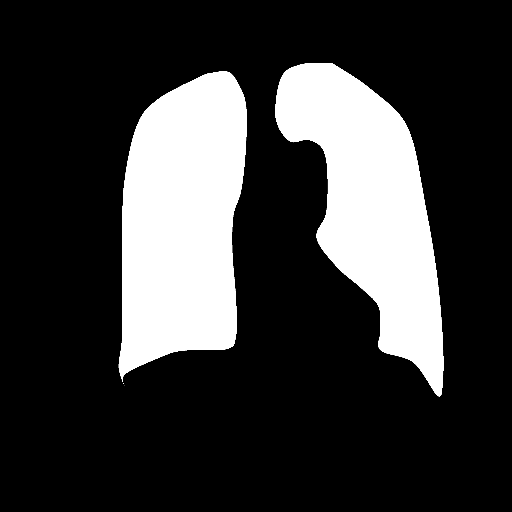}
	\end{minipage}

	\caption{Examples of Tuberculosis Chest X-rays (Shenzhen)dataset}
	\label{fig:fig5}
\end{figure}

\begin{itemize}
\item \textbf{Tuberculosis Chest X-rays dataset} \cite{jaeger2014two}. The dataset was acquired from the Department of Health and Human Services, Montgomery County, Maryland, USA, and Shenzhen No. 3 People’s Hospital in China. The chest X-rays are from outpatient clinics and were captured as part of the daily hospital routine within one month, mostly in September 2012, using a Philips DR Digital Diagnose system. The dataset contains normal cases and cases with TB manifestations. In the experiments of this paper, we do not need to consider whether the samples have TB manifestation but focus on the segmentation effect of the model on the lung region. We selected 566 clear chest x-ray images to form the experimental dataset and randomly divided the dataset into 452 training samples and 114 test samples in a ratio of 4:1. Some of the images and labels are shown in Fig. \ref{fig:fig5}.
\end{itemize}

\begin{figure}[htbp]
	\centering
	\begin{minipage}{0.19\linewidth}
		\centering
		\includegraphics[width=0.9\linewidth]{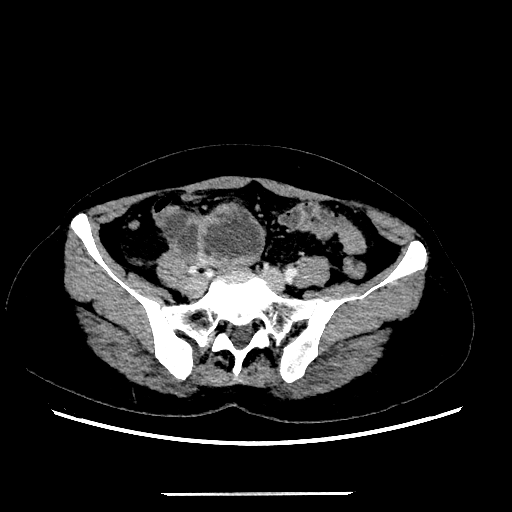}
	\end{minipage}
	\begin{minipage}{0.19\linewidth}
		\centering
		\includegraphics[width=0.9\linewidth]{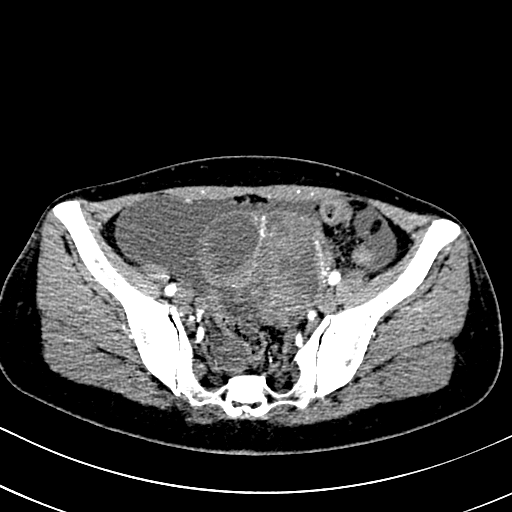}
	\end{minipage}
	\begin{minipage}{0.19\linewidth}
		\centering
		\includegraphics[width=0.9\linewidth]{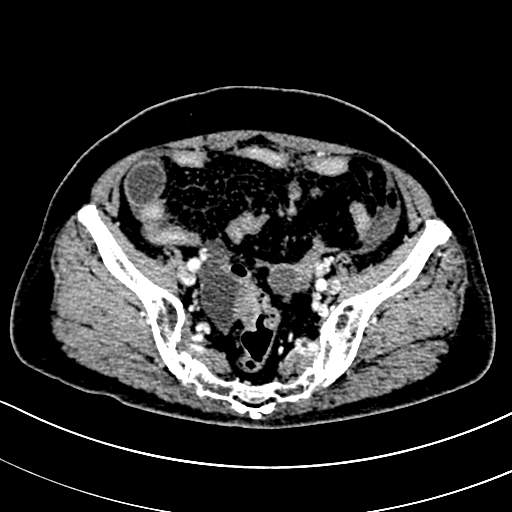}
	\end{minipage}
	\begin{minipage}{0.19\linewidth}
		\centering
		\includegraphics[width=0.9\linewidth]{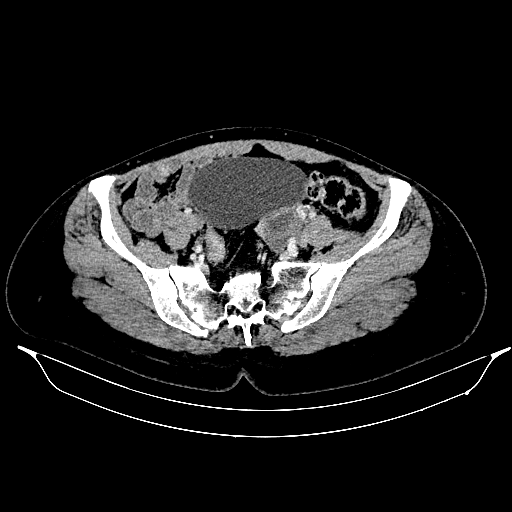}
	\end{minipage}
 
        \vspace{0.2cm}
        
	\begin{minipage}{0.19\linewidth}
		\centering
		\includegraphics[width=0.9\linewidth]{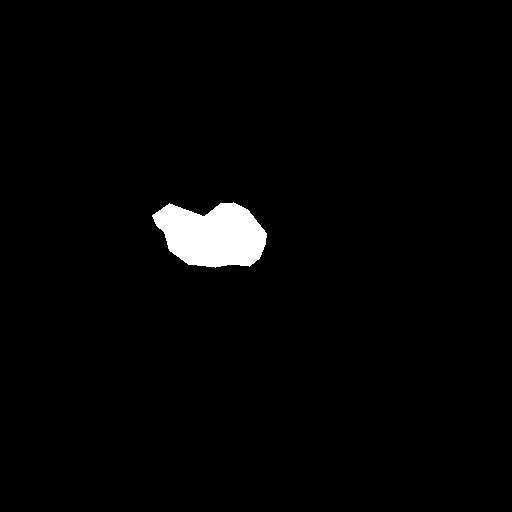}
	\end{minipage}
	\begin{minipage}{0.19\linewidth}
		\centering
		\includegraphics[width=0.9\linewidth]{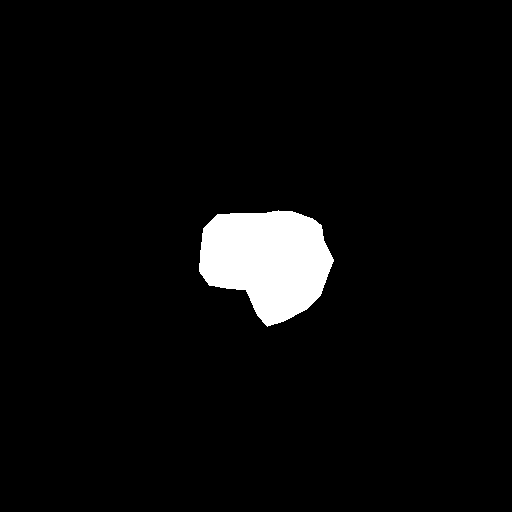}
	\end{minipage}
	\begin{minipage}{0.19\linewidth}
		\centering
		\includegraphics[width=0.9\linewidth]{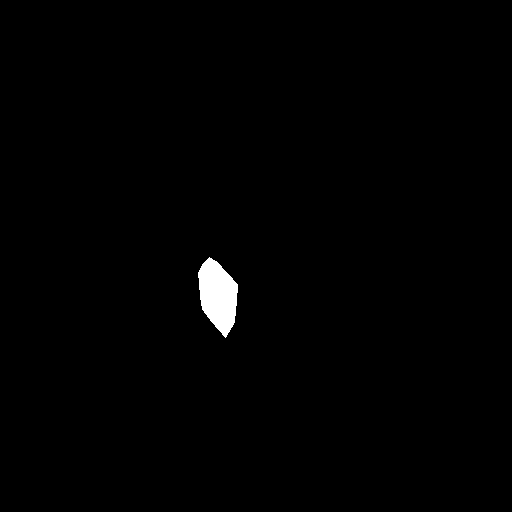}
	\end{minipage}
	\begin{minipage}{0.19\linewidth}
		\centering
		\includegraphics[width=0.9\linewidth]{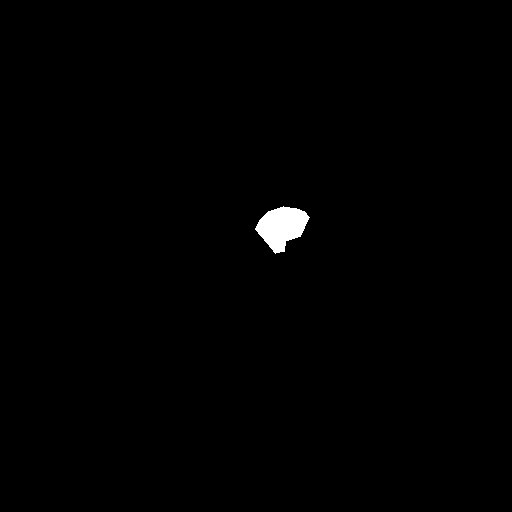}
	\end{minipage}
 
	\caption{Examples of Ovarian Tumors dataset}
	\label{fig:fig6}
\end{figure}

\begin{itemize}
\item \textbf{Ovarian Tumors dataset}. We collected a 2D CT segmentation dataset of abdominopelvic ovarian masses, which contains CT images of the patients as well as segmentation labels. The entire dataset was selected and annotated by physicians specializing in obstetrics and gynecology at the People's Hospital of Sichuan Province, China. The pelvic CT images of 123 patients were collected using Computed Tomography (CT), including 35 patients with benign tumors, 83 with malignant tumors, and 5 with unknown tumor benignity or malignancy. Since the CT images are three-dimensional, we need to slice the CT images of each patient horizontally and select several two-dimensional images containing the masses to construct the sample set. Based on this method, we constructed a CT image dataset containing 4050 abdominopelvic CT images of 123 patients. Next, we randomly selected 3092 CT images from 98 patients as the training set and 958 images from 25 patients as the test set. Such a construction method ensures that different images of the same patient will not appear in the training and test sets at the same time so that the segmentation effect of the model will not be affected by the image pattern of the same patient. Some images and labels are shown in Fig. \ref{fig:fig6}. 
\end{itemize}

\subsection{Implementation Details}

This subsection introduces the training parameters and loss function used in our experiments.

\begin{itemize}
\item \textbf{Loss function}. The loss functions used in our work are binary cross-entropy and Dice coefficient. Binary cross-entropy is utilized to evaluate the performance of the binary classification model, where we consider the regions of interest and non-interest in the images as 1 and 0, respectively. The formula for binary cross-entropy loss is described as Formula \eqref{equ1}, where y represents the actual labels of each pixel in the image and $\hat{y}$ is the predicted value of each pixel determined by the model. On the other hand, the Dice coefficient is employed to measure the overlap between the predicted image and the ground truth image. The formula of Dice loss is described as Formula \eqref{equ2}, in which TP, FP, TN and FN respectively represent the predicted true positive, false positive, true negative and false negative in the Confusion matrix, as shown in Table \ref{tab:tab1} in image segmentation. 
\end{itemize}

\begin{equation}
\begin{aligned}
L_{BCE}=-[(1-y)log(1-\hat{y})+ylog\hat{y}]
\end{aligned}
\label{equ1}
\end{equation}

\begin{equation}
\begin{aligned}
L_{DSC}=\ 1-\frac{2TP}{2TP+FP+FN}
\end{aligned}
\label{equ2}
\end{equation}

\begin{table}[htbp]
\caption{Confusion Matrix in image segmentation}
\centering
\begin{tabular}{cccc}
\hline
\multicolumn{2}{c}{\multirow{2}{*}{}} & \multicolumn{2}{c}{Predicted Sample} \\ \cmidrule{3-4} 
\multicolumn{2}{c}{}     &  Positive   & Negative \\ \hline
\multirow{2}{*}{\makecell{Actual \\ Sample}}   & Positive 
& \makecell{TP \\ (Pixels belong to real regions \\ of interest and are predicted \\ as regions of interest)}  
& \makecell{FP \\ (Pixels belong to real regions \\ of interest but are predicted \\ as regions of no interest)} \\ \cmidrule{2-4} 
&  Negative &  \makecell{FN\\(Pixels belong to real regions \\ of no interest and are predicted \\ as regions of interest)} 
& \makecell{TN\\(Pixels belong to real regions \\ of no interest and are predicted \\ as regions of no interest)}  \\ \hline
\end{tabular}
\label{tab:tab1}
\end{table}

\begin{itemize}
\item \textbf{Training parameters}. The experiments are trained and tested based on Python and Pytorch implementations. For the fully convolutional models, U-Net and UNet++, the input image size is set to 512x512. The models are trained using the Adam optimizer with a learning rate of 1e-3 and weight decay of 1e-8 for backpropagation. For the TransUNet model, the input image size and patch size are set to 512x512 and 16, respectively. The Transformer backbone and ResNet-50 utilize pre-trained weights from ImageNet. The model is optimized using the SGD optimizer with a momentum of 0.9 and weight decay of 1e-4 for backpropagation. For the Swin-Unet model, the input image size and patch size are set to 224x224 and 4, respectively. Pre-trained weights from ImageNet, provided by the original authors, are used for initializing the model parameters. The model is optimized using the SGD optimizer with a momentum of 0.9 and weight decay of 1e-4 for backpropagation. Table \ref{tab:tab2} shows the parameter settings for each model. Additionally, early stopping is introduced in the training of all models to prevent overfitting. 
\end{itemize}

\begin{table}[htbp]
\caption{parameter settings for each model}
\centering
\begin{tabular}{ccccc}
\hline
           & Input\_Size & Patch\_Size & Learning\_Rate & Optimizer \\ \cmidrule{2-5} 
U-Net      & 512*512     & 1           & 1e-3           & Adam      \\ \hline
UNet++     & 512*512     & 1           & 1e-3           & Adam      \\ \hline
Trans-UNet & 512*512     & 16          & 1e-2           & SGD       \\ \hline
Swin-UNet  & 224*224     & 4           & 1e-2           & SGD       \\ \hline
\end{tabular}
\label{tab:tab2}
\end{table}

\section{Experimental Results}
\subsection{Evaluation Metrics}

We employ the following medical image segmentation evaluation metrics to assess the performance of the model. Additionally, we observe the actual segmentation results to comprehensively evaluate the segmentation capability of the model. 

\begin{itemize}
\item \textbf{Dice}: The segmentation capability of a segmentation model is commonly measured using the Dice coefficient, which represents the similarity between two samples. It has a value range of [0, 1], where a higher value indicates better model performance. The Dice loss is described by Formula \eqref{equ2}.
\end{itemize}

\begin{itemize}
\item \textbf{HD95}: HD95 refers to the 95$\%$ Hausdorff distance, which quantifies the maximum distance between two sets at the 95th percentile. A smaller value indicates a higher similarity between the two sets. The HD95 is described by Formula \eqref{equ3}, where h(A, B) denotes the maximum value in the shortest distance to each pixel in B calculated for each pixel in A.
\end{itemize}

\begin{align}
\begin{aligned}
HD=\ {max}_{k95\%}(h(A,B),h(B,A)) \\
h(A,B)\ =\ max(a\in A)min(b\in B)||a-b|| \\
h(B,A)\ =\ max(b\in B)min(a\in A)||a-b|| \\
\end{aligned}
\label{equ3}
\end{align}

\begin{itemize}
\item \textbf{IoU}: The IoU (Intersection over Union) score is a standard performance measure for image segmentation problems. It also measures the similarity between two samples. The IoU score is described by Formula \eqref{equ4}.
\end{itemize}

\begin{equation}
\begin{aligned}
IoU=\frac{TP}{TP+FP+FN}
\end{aligned}
\label{equ4}
\end{equation}

\begin{itemize}
\item \textbf{Accuracy}: Accuracy is one of the commonly used metrics to measure the segmentation capability of a model. It specifically calculates the ratio of correctly classified pixels to the total number of pixels in the dataset. The accuracy is described by Formula \eqref{equ5}.
\end{itemize}

\begin{equation}
\begin{aligned}
Acc=\frac{TP+TN}{TP+FN+FP+TN}
\end{aligned}
\label{equ5}
\end{equation}

\begin{itemize}
\item \textbf{Precision}: Precision, also known as the positive predictive value, represents the ratio of true positive samples to the total predicted positive samples by the model. A higher precision value, closer to 1, is desired. Precision is described by Formula \eqref{equ6}.
\end{itemize}

\begin{equation}
\begin{aligned}
Precision=\frac{TP}{TP+FP}
\end{aligned}
\label{equ6}
\end{equation}

\begin{itemize}
\item \textbf{Recall}: Recall, also known as sensitivity or true positive rate, represents the ratio of true positive samples to the total actual positive samples. A higher recall value, closer to 1, is desirable. Recall is described by Formula \eqref{equ7}.
\end{itemize}

\begin{equation}
\begin{aligned}
Recall=\frac{TP}{TP+FN}
\end{aligned}
\label{equ7}
\end{equation}

\subsection{Experimental Results}

\begin{figure}[htbp]
	\centering
        \begin{minipage}{0.16\linewidth}
        \centering
            Images
        \end{minipage}
        \begin{minipage}{0.16\linewidth}
        \centering
            U-Net
        \end{minipage}
        \begin{minipage}{0.16\linewidth}
        \centering
            UNet++
        \end{minipage}
        \begin{minipage}{0.16\linewidth}
        \centering
            TransUnet
        \end{minipage}
        \begin{minipage}{0.16\linewidth}
        \centering
            Swin-Unet
        \end{minipage}
        \begin{minipage}{0.16\linewidth}
        \centering
            LABEL
        \end{minipage}
        \vspace{0.2cm}
	\begin{minipage}{0.16\linewidth}
		\centering
		\includegraphics[width=0.9\linewidth]{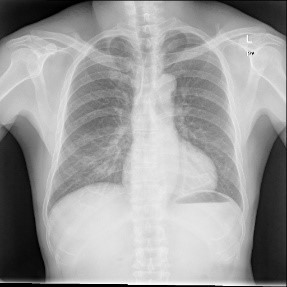}
	\end{minipage}
	\begin{minipage}{0.16\linewidth}
		\centering
		\includegraphics[width=0.9\linewidth]{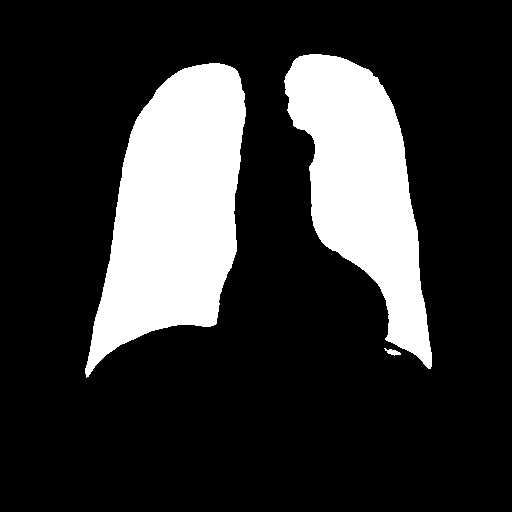}
	\end{minipage}
	\begin{minipage}{0.16\linewidth}
		\centering
		\includegraphics[width=0.9\linewidth]{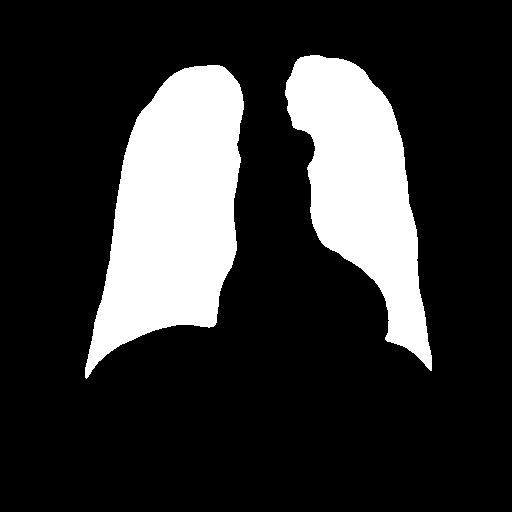}
	\end{minipage}
	\begin{minipage}{0.16\linewidth}
		\centering
		\includegraphics[width=0.9\linewidth]{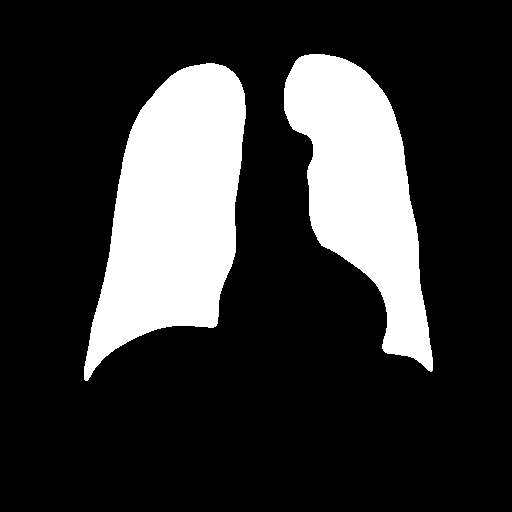}
	\end{minipage}
	\begin{minipage}{0.16\linewidth}
		\centering
		\includegraphics[width=0.9\linewidth]{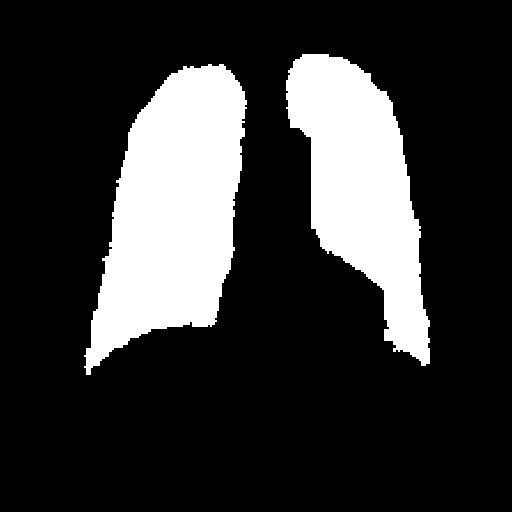}
	\end{minipage}
    \begin{minipage}{0.16\linewidth}
		\centering
		\includegraphics[width=0.9\linewidth]{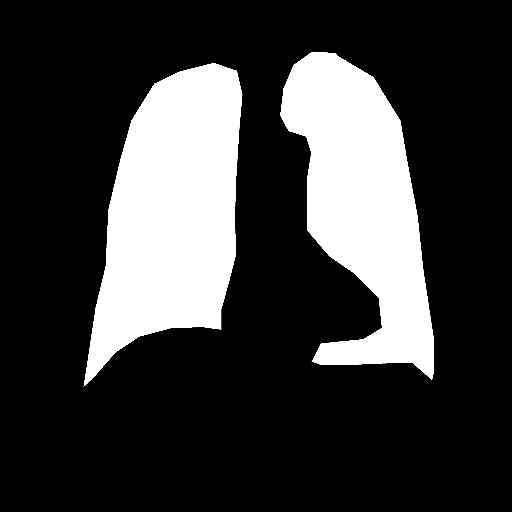}
	\end{minipage}
        \vspace{0.2cm}
	\begin{minipage}{0.16\linewidth}
		\centering
		\includegraphics[width=0.9\linewidth]{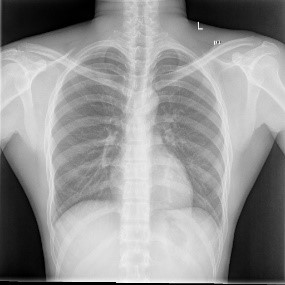}
	\end{minipage}
	\begin{minipage}{0.16\linewidth}
		\centering
		\includegraphics[width=0.9\linewidth]{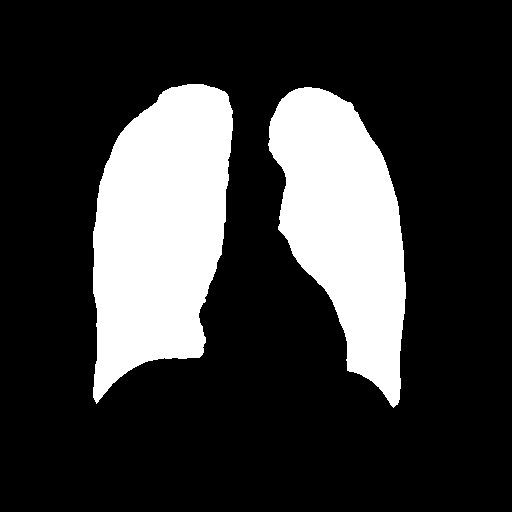}
	\end{minipage}
	\begin{minipage}{0.16\linewidth}
		\centering
		\includegraphics[width=0.9\linewidth]{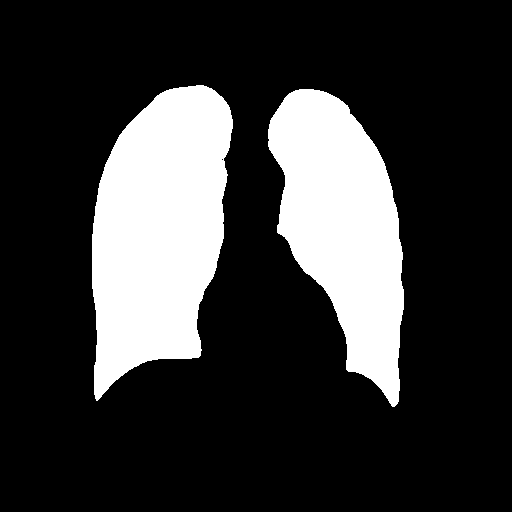}
	\end{minipage}
        \begin{minipage}{0.16\linewidth}
		\centering
		\includegraphics[width=0.9\linewidth]{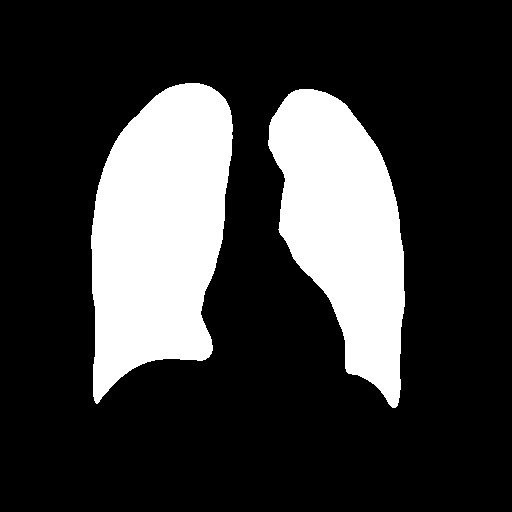}
	\end{minipage}
        \begin{minipage}{0.16\linewidth}
		\centering
		\includegraphics[width=0.9\linewidth]{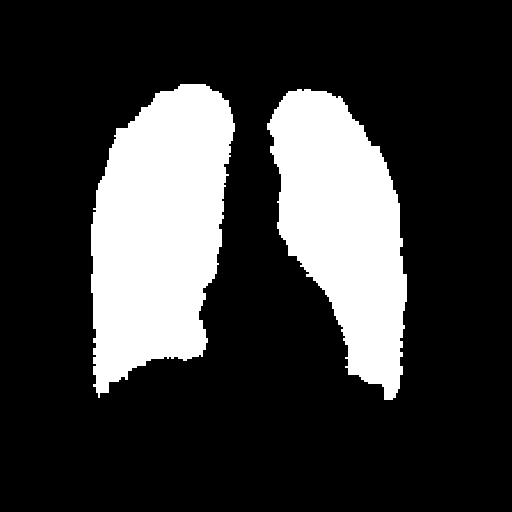}
	\end{minipage}
        \begin{minipage}{0.16\linewidth}
		\centering
		\includegraphics[width=0.9\linewidth]{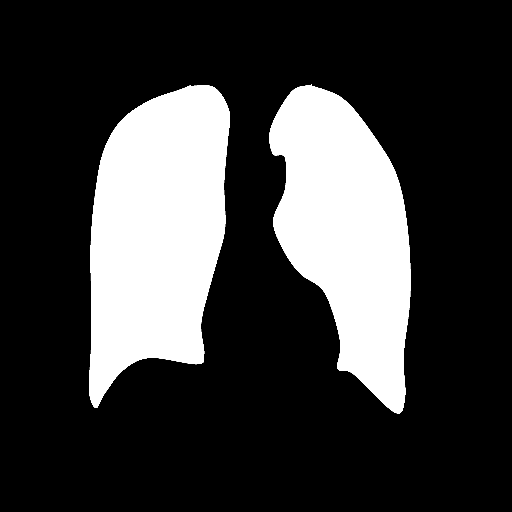}
	\end{minipage}

	\caption{Tuberculosis Chest X-rays (Shenzhen)dataset segmentation results}
	\label{fig:fig7}
\end{figure}

We first evaluate the performance of each model on the publicly available dataset, the Tuberculosis Chest X-rays dataset. Table \ref{tab:tab3} presents the experimental results of each model, while Fig. \ref{fig:fig7} provides a visual representation of the model's performance. The results indicate that the TransUNet model achieves the best performance across all six metrics, with 96.45\% (DSC↑), 10.75 (HD↓), 93.25\% (IoU↑), 98.16\% (Acc↑), 97.36\% (Precision↑), and 95.72\% (Recall↑). Moreover, all models demonstrate excellent performance in the lung segmentation task, with mIoU (mean Intersection over Union) values exceeding 91\% for all four segmentation methods. Excluding samples with less unclear sampling from the test set, the remaining samples achieve DSC scores of over 82\%. According to their segmentation results, all four methods can effectively meet the segmentation requirements.

\begin{table}[htbp]
\caption{Tuberculosis Chest X-rays (Shenzhen) dataset test index}
\centering
\begin{tabular}{ccccccc}
\hline
           & DSC↑    & HD95↓ & IoU↑    & Acc↑    & Precision↑ & Recall↑ \\ \hline
U-Net      & 95.32\% & 14.23 & 91.24\% & 97.69\% & 96.13\%    & 94.78\% \\ \hline
UNet++     & 95.83\% & 11.75 & 92.15\% & 97.95\% & 97.31\%    & 94.62\% \\ \hline
TransUNet & \color{red}{96.45}\% & \color{red}{10.75} & \color{red}{93.25}\% & \color{red}{98.16}\% & \color{red}{97.36}\%    & \color{red}{95.72}\% \\ \hline
Swin-Unet  & 95.71\% & 12.10 & 91.88\% & 97.80\% & 96.81\%    & 94.79\% \\ \hline
\end{tabular}
\label{tab:tab3}
\end{table}

\begin{figure}[htbp]
	\centering
        \begin{minipage}{0.16\linewidth}
        \centering
            Images
        \end{minipage}
        \begin{minipage}{0.16\linewidth}
        \centering
            U-Net
        \end{minipage}
        \begin{minipage}{0.16\linewidth}
        \centering
            UNet++
        \end{minipage}
        \begin{minipage}{0.16\linewidth}
        \centering
            TransUnet
        \end{minipage}
        \begin{minipage}{0.16\linewidth}
        \centering
            Swin-Unet
        \end{minipage}
        \begin{minipage}{0.16\linewidth}
        \centering
            LABEL
        \end{minipage}
        \vspace{0.2cm}
	\begin{minipage}{0.16\linewidth}
		\centering
		\includegraphics[width=0.9\linewidth]{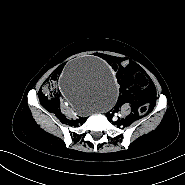}
	\end{minipage}
	\begin{minipage}{0.16\linewidth}
		\centering
		\includegraphics[width=0.9\linewidth]{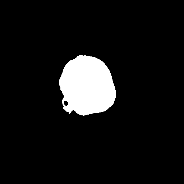}
	\end{minipage}
	\begin{minipage}{0.16\linewidth}
		\centering
		\includegraphics[width=0.9\linewidth]{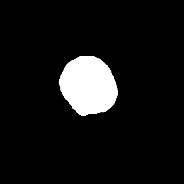}
	\end{minipage}
	\begin{minipage}{0.16\linewidth}
		\centering
		\includegraphics[width=0.9\linewidth]{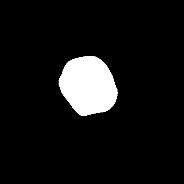}
	\end{minipage}
	\begin{minipage}{0.16\linewidth}
		\centering
		\includegraphics[width=0.9\linewidth]{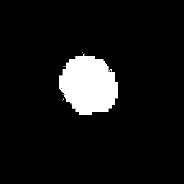}
	\end{minipage}
    \begin{minipage}{0.16\linewidth}
		\centering
		\includegraphics[width=0.9\linewidth]{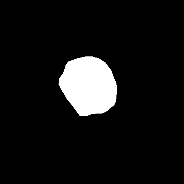}
	\end{minipage}
        \vspace{0.2cm}
	\begin{minipage}{0.16\linewidth}
		\centering
		\includegraphics[width=0.9\linewidth]{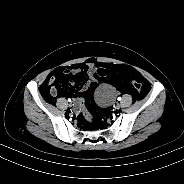}
	\end{minipage}
	\begin{minipage}{0.16\linewidth}
		\centering
		\includegraphics[width=0.9\linewidth]{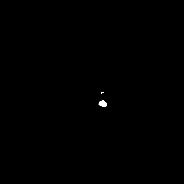}
	\end{minipage}
        \begin{minipage}{0.16\linewidth}
		\centering
		\includegraphics[width=0.9\linewidth]{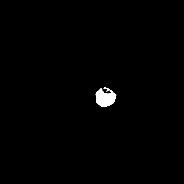}
	\end{minipage}
        \begin{minipage}{0.16\linewidth}
		\centering
		\includegraphics[width=0.9\linewidth]{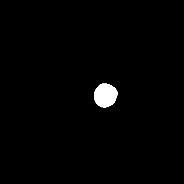}
	\end{minipage}
        \begin{minipage}{0.16\linewidth}
		\centering
		\includegraphics[width=0.9\linewidth]{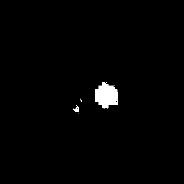}
	\end{minipage}
        \begin{minipage}{0.16\linewidth}
		\centering
		\includegraphics[width=0.9\linewidth]{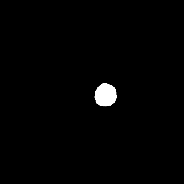}
	\end{minipage}
        \vspace{0.2cm}
        \begin{minipage}{0.16\linewidth}
		\centering
		\includegraphics[width=0.9\linewidth]{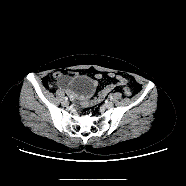}
	\end{minipage}
	\begin{minipage}{0.16\linewidth}
		\centering
		\includegraphics[width=0.9\linewidth]{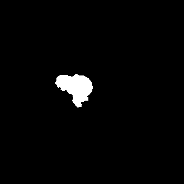}
	\end{minipage}
        \begin{minipage}{0.16\linewidth}
		\centering
		\includegraphics[width=0.9\linewidth]{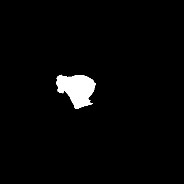}
	\end{minipage}
        \begin{minipage}{0.16\linewidth}
		\centering
		\includegraphics[width=0.9\linewidth]{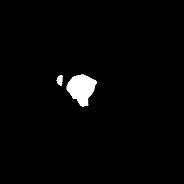}
	\end{minipage}
        \begin{minipage}{0.16\linewidth}
		\centering
		\includegraphics[width=0.9\linewidth]{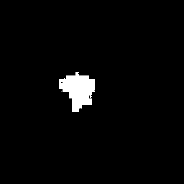}
	\end{minipage}
        \begin{minipage}{0.16\linewidth}
		\centering
		\includegraphics[width=0.9\linewidth]{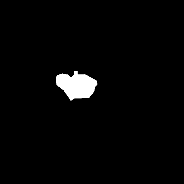}
	\end{minipage}
        \vspace{0.2cm}
        \begin{minipage}{0.16\linewidth}
		\centering
		\includegraphics[width=0.9\linewidth]{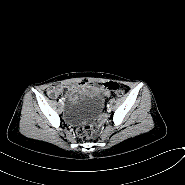}
	\end{minipage}
	\begin{minipage}{0.16\linewidth}
		\centering
		\includegraphics[width=0.9\linewidth]{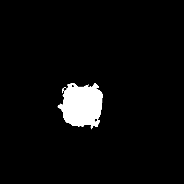}
	\end{minipage}
        \begin{minipage}{0.16\linewidth}
		\centering
		\includegraphics[width=0.9\linewidth]{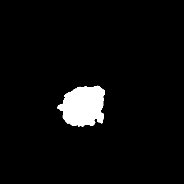}
	\end{minipage}
        \begin{minipage}{0.16\linewidth}
		\centering
		\includegraphics[width=0.9\linewidth]{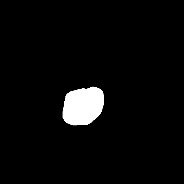}
	\end{minipage}
        \begin{minipage}{0.16\linewidth}
		\centering
		\includegraphics[width=0.9\linewidth]{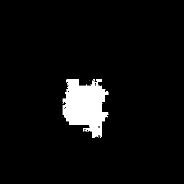}
	\end{minipage}
        \begin{minipage}{0.16\linewidth}
		\centering
		\includegraphics[width=0.9\linewidth]{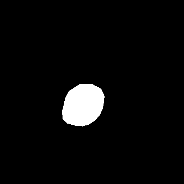}
	\end{minipage}
        \vspace{0.2cm}
        \begin{minipage}{0.16\linewidth}
		\centering
		\includegraphics[width=0.9\linewidth]{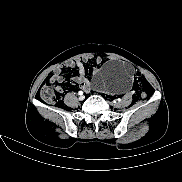}
	\end{minipage}
	\begin{minipage}{0.16\linewidth}
		\centering
		\includegraphics[width=0.9\linewidth]{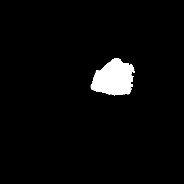}
	\end{minipage}
        \begin{minipage}{0.16\linewidth}
		\centering
		\includegraphics[width=0.9\linewidth]{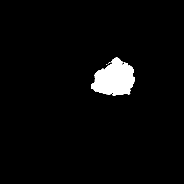}
	\end{minipage}
        \begin{minipage}{0.16\linewidth}
		\centering
		\includegraphics[width=0.9\linewidth]{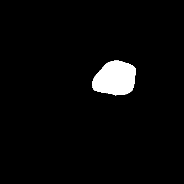}
	\end{minipage}
        \begin{minipage}{0.16\linewidth}
		\centering
		\includegraphics[width=0.9\linewidth]{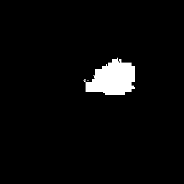}
	\end{minipage}
        \begin{minipage}{0.16\linewidth}
		\centering
		\includegraphics[width=0.9\linewidth]{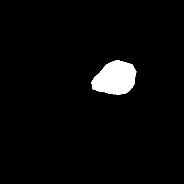}
	\end{minipage}

	\caption{Ovarian Tumors dataset segmentation results}
	\label{fig:fig8}
\end{figure}

We further evaluate the performance of each model on our dataset with different imaging modalities. Table \ref{tab:tab4} presents the results for the Ovarian Tumors dataset. The models are evaluated based on the average Dice coefficient and mIoU to assess the experimental results. The TransUNet model performs the best, followed by Swin-Unet, UNet++, and U-Net. When evaluating the results based on the average Hausdorff distance, the TransUNet model also performs best, followed by U-Net, Swin-Unet, and UNet++. In terms of average accuracy, the TransUNet model outperforms the others, followed by Swin-Unet, U-Net, and UNet++. When evaluating based on average precision, the TransUNet model performed the best, followed by U-Net, Swin-Unet, and UNet++. Notably, the Trans-UNet model demonstrates significantly higher precision than other models, indicating its strong ability to identify mass regions accurately. For average Recall, the Swin-Unet model performs the best, followed by TransUNet, UNet++, and U-Net. Fig. \ref{fig:fig8} provides a visual representation of the model’s performance. Considering all the metrics and actual segmentation results, the Trans-UNet model achieves the best segmentation performance, followed by Swin-Unet, U-Net, and UNet++. The TransUNet model demonstrates the highest performance across all five evaluation metrics, with 89.18\% (DSC↑), 22.35 (HD↓), 82.73\% (IoU↑), 99.02\% (ACC↑), and 92.28\% (Recall↑). Its predicted results are largely similar to the ground truth labels. On the other hand, Swin-Unet, UNet++, and U-Net, which exhibit lower evaluation metrics, do not produce perfect results in the actual predictions. 

\begin{table}[htbp]
\caption{Ovarian Tumors dataset test index}
\centering
\begin{tabular}{ccccccc}
\hline
           & DSC↑    & HD95↓ & IoU↑    & Acc↑    & Precision↑ & Recall↑ \\ \hline
U-Net      & 79.12\% & 27.63 & 70.07\% & 96.88\% & 80.93\%    & 82.21\% \\ \hline
UNet++     & 79.87\% & 34.43 & 71.74\% & 94.48\% & 77.43\%    & 87.01\% \\ \hline
Trans-UNet & \color{red}{89.18}\% & \color{red}{22.35} & \color{red}{82.73}\% & \color{red}{99.02}\% & \color{red}{92.28}\%    & 89.20\% \\ \hline
Swin-UNet  & 83.06\% & 30.80 & 73.39\% & 98.22\% & 78.98\%    & \color{red}{91.41}\% \\ \hline
\end{tabular}
\label{tab:tab4}
\end{table}

We further examined the segmentation results of each model when the Dice coefficient was extremely low (less than 20\%). We considered these results as completely unacceptable segmentation outcomes. The statistical count of such results is shown in Table \ref{tab:tab5}. It indicates that after introducing the Transformer module, the models were able to capture global information and thus exhibited fewer instances of completely erroneous judgments regarding the main region of the ovarian mass. 

\begin{table}[htbp]
\caption{Statistics of the number of samples with dice coefficient less than 20\%}
\centering
\begin{tabular}{cccc}
\hline
U-Net & UNet++ & Trans-UNet & Swin-UNet \\ \hline
34    & 56     & 5          & 5         \\ \hline
\end{tabular}
\label{tab:tab5}
\end{table}

\subsection{Discussions}

In our experiments, all models employed the U-shaped architecture, and the introduction of the U-Net model was of significant importance in medical image segmentation. U-Net combines an encoder and a decoder, allowing for accurate segmentation by utilizing information at different scales while preserving high-resolution features. This design enables U-Net to achieve more accurate results in medical image segmentation tasks, improving the localization and segmentation precision of lesions. Additionally, U-Net exhibits good scalability and can be improved by adding or adjusting network layers, modifying the network structure, and more. This flexibility enables the U-Net model to be applied to various medical image segmentation tasks and integrated with and optimized alongside other deep learning models.

Initially, Transformer models are not considered promising for medical image segmentation due to their inherent lack of localization ability. However, TransUNet introduces a structure that combines Transformers with convolutional neural networks, forming an effective encoder and improving segmentation performance. On the other hand, if convolutional networks are not combined with Transformers, the final results are not ideal, as demonstrated by Swin-Unet. 

As for the dataset, the segmentation of ovarian masses is highly challenging, and the segmented regions of the images often differ in size, shape, location, and texture, making detecting masses more difficult. In the lung segmentation dataset, identifying the lung extent is easy because it has distinct features compared to the background, and each model performs excellently in the lung segmentation task. 

\section{Dataset Challenges and Issues}

The biggest challenge in applying supervised learning in medical image processing is medical image labeling. Supervised learning requires the input of a large number of annotated samples to obtain good performance and stable generalization. However, collecting such a large dataset of annotated cases is usually a very daunting task. Medical images require the interpretation of professional clinicians to collect, label, and annotate medical images. Another important problem in medical image processing is data imbalance. In unbalanced datasets, the class distribution between categories is asymmetric; for example, there is a natural imbalance in the number of benign and malignant patients in the ovarian mass dataset (the number of abnormal patients is larger than that of normal patients). Unbalanced data can significantly affect model performance.

Several methods have been widely used to solve the above problems. Data augmentation can increase the number of training datasets and balance the ratio of positive and negative samples by applying a set of affine transformations to the samples, such as flipping, rotating, mirroring \cite{milletari2017hough}, and enhancing the color (gray) values \cite{golan2016lung}. Migrating learning from successful models implemented in the same or other domains is another solution to the above problem. Compared to data enhancement, migration learning is a more specific solution that requires only modest computational resources and a smaller amount of labeled data to significantly reduce the error rate in medical image segmentation \cite{beevi2019automatic}. 

\section{Conclusions}

In this paper, we first give a general introduction to medical segmentation and then investigate the four most representative medical image segmentation models, i.e., U-Net, UNet++, TransUNet, and Swin-Unet. Moreover, quantitative performance evaluations of these models are conducted on two benchmark datasets. Also, to help researchers in related fields understand these models quickly and to model new segmentation tasks, we share all the experimental source code on GitHub and the detailed parameters of the model setup. 

In recent years, image segmentation based on machine learning has developed rapidly. Meta \cite{kirillov2023segment} proposed the segmentation everything model (SAM), revolutionizing image segmentation. It is the first time to introduce the concept of foundation models in image segmentation with zero-shot migration. Unlike previous image segmentation models that can only handle a particular class of images, SAM can handle all images and achieve accurate image segmentation by the prompt. Ma et al. \cite{ma2023segment} proposed MedSAM for general-purpose image segmentation, which is the first attempt to apply SAM to the medical field, and it outperforms the default SAM model on medical segmentation tasks. Large model-based image segmentation has become a future trend in image segmentation, which is a promising research direction with a wide range of prospects. In future work, we will combine large model-based image segmentation with medical images, expecting to contribute to medical image analysis by solving the problem of difficult medical image acquisition through a segmentation model with zero-shot migration.

\section{Acknowledgements}

This research was supported by Open Project of Network and Data Security Key Laboratory of Sichuan Province(NSD2021-6), Clinical Research and Transformation Fund of Sichuan Provincial People's Hospital(2021LY24), and the Key Research Project of Science and Technology of Sichuan Province(2022YFS0087)(2023YFS0039).

\bibliography{AROMISM}

\end{document}